\documentclass[preprint]{revtex4}

\usepackage{amsmath}
\usepackage{mathrsfs}
\usepackage{comment}
\usepackage{amssymb}
\usepackage{graphicx}
\usepackage[colorlinks,
            linkcolor=blue,
            anchorcolor=blue,
            citecolor=blue]{hyperref}
\usepackage{subfigure}
\newtheorem{theorem}{Theorem}

\def\ra{\rangle}
\def\la{\langle}
\allowdisplaybreaks[4]

\begin{document}
\title{Parameterized Multi-observable Sum Uncertainty Relations}
\author{Jing-Feng Wu$^{1}$}
\author{Qing-Hua Zhang$^{1,}$\footnotemark[1]}
\author{Shao-Ming Fei$^{1,2,}$\footnotemark[1]}

\affiliation{$^1$School of Mathematical Sciences, Capital Normal University, 100048
Beijing, China\\
$^2$Max-Planck-Institute for Mathematics in the Sciences, 04103 Leipzig, Germany}

\renewcommand{\thefootnote}{\fnsymbol{footnote}}

\footnotetext[1]{Corresponding authors. \\
\href{mailto:2190501022@cnu.edu.cn}{2190501022@cnu.edu.cn(Q. H. Zhang)}.\\
\href{mailto:feishm@cnu.edu.cn}{feishm@cnu.edu.cn(S. M. Fei)}.}
\bigskip

\bigskip

\begin{abstract}
The uncertainty principle is one of the fundamental features of quantum mechanics and plays an essential role in quantum information theory. We study uncertainty relations based on variance for arbitrary finite $N$ quantum observables. We establish a series of parameterized uncertainty relations in terms of the parameterized norm inequalities, which improve the exiting variance-based uncertainty relations. The lower bounds of our uncertainty inequalities are non-zero unless the measured state is a common eigenvector of all the observables. Detailed examples are provided to illustrate the tightness of our uncertainty relations.
\end{abstract}

\maketitle

\section{Introduction}\label{Sec1}
The uncertainty principle is one of the cornerstones of quantum mechanics, which reveals the intrinsic differences between classical and quantum world. The uncertainty principle was firstly introduced by Heisenberg in 1927 \cite{Heisenberg1927}. It shows that if one measures the momentum of a particle with certainty, one can not measure its position with certainty at the same time. Since then a lot of research works have been dedicated to interpret the uncertainty relations via different forms, such as in terms of quantum variance \cite{Nielsen2002Quantum,PhysRev.34.163,schrodinger1930sitzungsberichte,PhysRevLett.113.260401,Kennard1927Zur,Schrodinger1930Zum,Robertson1929The,mondal2017tighter,PhysRevResearch.4.013076,PhysRevResearch.4.013075}, entropy \cite{Maassen1988PhysRevLett.60.1103,Wu2009PhysRevA.79.022104,Coles2017RevModPhys.89.015002}, noise and disturbance \cite{buschPhysRevLett.111.160405}, successive measurement \cite{DeutschPhysRevLett.50.631,DistlerPhysRevA.87.062112}, majorization technique \cite{Pucha_a_2013,friedland2013universal}, skew information \cite{luoPhysRevLett.91.180403,Zhang_2021note,zhang2021tighter,ma2022product} etc. These uncertainty relations play an important role in a wide range of applications such as entanglement detection \cite{PhysRevLett.92.117903, zhang2020sufficient,zhang2021multipartite}, quantum metrology \cite{PhysRevLett.96.010401}, quantum steering \cite{SchneelochPhysRevA.87.062103}, quantum gravity \cite{hall2005exact} and quantum cryptography \cite{Fuchs1996Quantum}.

Robertson \cite{PhysRev.34.163} generalized the variance-based uncertainty relation for position and momentum to any two observables $A$ and $B$,
\begin{equation}\label{Eq1}
  \Delta A  \Delta B \geq \frac{1}{2} |\la \psi | [A, B] | \psi \ra |,
\end{equation}
where $\Delta$ stands for the standard deviation of the observable with respect to a fixed state $|\psi\ra$ and $[A, B]$ represents the commutator of the observables $A$ and $B$. Eq.~(\ref{Eq1}) was later improved by Schr\"{o}dinger \cite{schrodinger1930sitzungsberichte},
\begin{equation}\label{schrodinger}
\Delta A\Delta B \geq \frac{1}{2}\sqrt{|\langle\{A, B\}\rangle-\langle A\rangle\langle B\rangle |^{2}+|\langle[A, B]\rangle |^{2}}.
\end{equation}
Here, the lower bounds of the inequalities (\ref{Eq1}) and (\ref{schrodinger}) may vanish even if the observables $A$ and $B$ are not commutative. For instance, when the measured state $|\psi\ra$ is an eigenvector of either $A$ or $B$, the right hands of the inequality (\ref{Eq1}) and (\ref{schrodinger}) are trivially zero. To overcome the flaw, uncertainty relations with respect to the sum of variances have been presented by Maccone and Pati \cite{PhysRevLett.113.260401},
\begin{align}
\label{Pati1}  \Delta^2  A + \Delta^2 B &\geq \pm i \la \psi | [A, B] | \psi \ra + | \la \psi | A \pm i B | \psi^{\perp} \ra |^2,\\
\label{Pati2}                  \Delta^2  &A + \Delta^2 B \geq \frac{1}{2} \Delta^2(A + B),
\end{align}
where the signs $\pm$ on the right-hand side of (\ref{Pati1}) are taken so that the $\pm i\la \psi | [A, B] | \psi \ra$ is positive, $|\psi^{\perp}\ra$ satisfies $\la \psi| \psi^{\perp} \ra = 0$. The lower bound in (\ref{Pati1}) is nonzero for most choices of $|\psi^{\perp}\ra$ if $A$ and $B$ are not commutative.

Besides the variance-based uncertainty relations with respect to pairs of non-commutative observables, the uncertainty relations related to three non-commutative observables such as the three components of spins and angular momentums \cite{kechrimparis2014heisenberg,dammeier2015uncertainty,ma2017experimental} have been also investigated. The variance-based uncertainty relations for general multiple observables have been further studied either in summation form \cite{chen2016variance, chen2015sum, chen2019tight} or in product form \cite{qin2016multi,xiao2016mutually}. For instance, Song $et\ al.$ derived an elegant variance-based uncertainty relation in \cite{song2017stronger},
\begin{equation}\label{Song_Eq}
\sum_{i = 1}^N \Delta_{\rho}^2(A_i) \geq \frac{1}{N} \Delta_{\rho}^2  (\sum_{i = 1}^N A_i  )
+ \frac{2}{N^2(N - 1)}  \Big [ \sum_{1 \leq i <j \leq N} \Delta_{\rho}(A_i - A_j)  \Big ]^2.
\end{equation}
Recently, based on the inequalities of vector norm, Zhang $et\ al.$ \cite{zhang2022note} proposed an improved variance-based sum uncertainty
relation for $N$ arbitrary incompatible observables,
\begin{equation}\label{Zhang_Eq} 
\begin{aligned}
\sum_{i = 1}^N \Delta_{\rho}^2(A_i) \geq \max_{x \in \{0, 1\}}
&\frac{1}{2 N - 2}  \bigg\{ \frac{2}{N(N - 1)}  \Big [ \sum_{1 \leq i <j \leq N} \Delta_{\rho}(A_i + (- 1)^x A_j)  \Big ]^2 \\
&+ \sum_{1 \leq i < j \leq N} \Delta_{\rho}^2(A_i + (- 1)^{x + 1} A_j )  \bigg\}.
\end{aligned}
\end{equation}

This paper is aimed to improve these uncertainty relations for $N$ arbitrary observables. Motivated by the skew information-based uncertainty relations proposed in \cite{zhang2022note} and \cite{li2022metric}, we combine the parameterized parallelogram law of vector norm with Cauchy-Schwarz inequality to improve the lower bounds of uncertainty relations for $N$ observables.


\section{uncertainty relations via variance}\label{Sec2}
Denote $H_d$ the Hilbert space with $d$ dimension. Let $G=(g_{ij})_{l\times p}$ be a rectangular matrix with entries $g_{ij}$. The vectorization of $G$ is given by the vector $|G\rangle=(g_{11},\dots,g_{l1},\dots,g_{1p},\dots,g_{lp})^t$ with $t$ denoting the transpose. It is verified that $|GT\ra=(I\otimes G)|T\ra$ for any matrix $T$ and identity $I$ in suitable size. The quantum variance of any quantum state $\rho$ on $H_d$ with respect to an observable $M$ is defined by
\begin{equation}
\begin{aligned}
\Delta^2_{\rho} (M)&={\rm Tr} (\rho M^2)-[{\rm Tr} (\rho M)]^2={\rm Tr} (\sqrt{\rho}(\delta M)^2\sqrt{\rho})\\
&=\la\delta M\sqrt{\rho}|\delta M\sqrt{\rho}\ra=\la \sqrt{\rho}| I_d\otimes (\delta M)^2 |\sqrt{\rho}\ra\\
&=\| I_d\otimes \delta M |\sqrt{\rho}\ra\|^2,
\end{aligned}
\end{equation}
where $I_d$ is the identity operator in $H_d$, $\delta M=M-{\rm Tr}(\rho M)$, $|\sqrt{\rho}\ra$ denotes the vectorization of $\sqrt{\rho}$ and
$\|\cdot\|$ the $2$-norm of a vector. Especially, when the quantum state $\rho$ is pure, that is $\rho=|\psi\rangle \langle\psi|$, one has $\Delta^2_{\rho} (M)=\|\delta M |\psi\rangle \|^2$. By using the parallelogram law of vector $2$-norm,
\begin{equation}
(2N-2) \sum_{i=1}^{N} \| a_i\|^2 = \sum_{1\leq i<j \leq N} \| a_i+a_j \|^2 + \sum_{1\leq i<j \leq N} \| a_i-a_j \|^2
\end{equation}
for a set of vectors $a_i$, $i=1,...,N$, we have the following uncertainty relation.

\begin{theorem} \label{Thm1}
For $N$ arbitrary observables $A_1, A_2, \dots, A_N$, the following  variance-based sum uncertainty relation holds for any quantum state $\rho$,
\begin{equation}\label{Thm1_Eq1}
\begin{aligned}
\sum_{i=1}^N \Delta_{\rho}^2 (A_i) \geq \rm{LB1}=\mathop{ \max_{ { x \in \{ 0, 1\} } \atop { y \in \{ 0, 1\} } } }
&\frac{1}{ (1 + \alpha^2) (N - 1) }  \bigg\{ \frac{2} { N (N -1) } \Big [ \sum_{ 1 \leq i < j \leq N } \Delta_{\rho} ( \alpha^{1 - x} A_i  + (- 1)^y \alpha^x A_j) \Big ]^2   \\
&  + \sum_{ 1 \leq i < j \leq N } \Delta_{\rho}^2 ( \alpha^ x A_i + (- 1)^{1 - y} \alpha^{1 - x} A_j) \bigg\},
\end{aligned}
\end{equation}
where $\alpha$ is any non-negative real number.
\end{theorem}

{\sf [Proof]} For all $x \in \{0, 1\}$ and $y\in \{0, 1\}$, the following parameterized parallelogram equality holds for $2$-norm of any vectors $a_i$,
\begin{equation}
\begin{aligned}
\sum_{ i = 1}^N \| a_i \|^2 = &\frac{1}{ ( 1 + \alpha^2 ) ( N - 1 ) }  \Big [ \sum_{ 1 \leq i < j \leq N } \| \alpha^{1 - x} a_i + (- 1)^y \alpha^x a_j \|^2  \\
&  + \sum_{ 1 \leq i < j \leq N } \| \alpha^x a_i + (- 1)^{1 - y} \alpha^{1 - x} a_j \|^2  \Big ].
\end{aligned}
\end{equation}
Using the Cauchy-Schwarz inequality,
\begin{equation}\label{cauchy}
\sum_{ 1 \leq i < j \leq N } \| \alpha^{1 - x} a_i + (- 1)^y \alpha^x a_j \|^2 \geq \frac{2}{ N ( N - 1 ) }  \Big [ \sum_{ 1 \leq i < j \leq N } \| \alpha^{1 - x} a_i + (- 1)^y \alpha^x a_j \|  \Big ]^2,
\end{equation}
we obtain
\begin{equation}
\begin{aligned}
\sum_{ i = 1 }^N \|a_i\|^2 \geq
&\frac{1}{ (1 + \alpha^2) (N - 1) }  \bigg\{ \frac{2} { N (N -1) }  \Big [ \sum_{ 1 \leq i < j \leq N } \| \alpha^{1 - x} a_i  + (- 1)^y \alpha^x a_j \|  \Big ]^2   \\
&  + \sum_{ 1 \leq i < j \leq N } \| \alpha^ x a_i + (- 1)^{1 - y} \alpha^{1 - x} a_j \|^2  \bigg\}.
\end{aligned}
\end{equation}
Set $\| a_i \|=\| I_d\otimes \delta A_i |\sqrt{\rho}\ra\|=\Delta_{\rho} A_i$ and $\| \alpha^{1 - x} a_i  + (- 1)^y \alpha^x a_j \|=\Delta_{\rho} ( \alpha^{1 - x} A_i  + (- 1)^y \alpha^x A_j) $. We have
\begin{equation}
\begin{aligned}
\sum_{ i = 1 }^N \Delta_{\rho}^2 (A_i) \geq
&\frac{1}{ (1 + \alpha^2) (N - 1) }  \bigg\{ \frac{2} { N (N -1) }  \Big [ \sum_{ 1 \leq i < j \leq N } \Delta_{\rho} ( \alpha^{1 - x} A_i  + (- 1)^y \alpha^x A_j)  \Big ]^2   \\
&  + \sum_{ 1 \leq i < j \leq N } \Delta_{\rho}^2 ( \alpha^ x A_i + (- 1)^{1 - y} \alpha^{1 - x} A_j)  \bigg\}.
\end{aligned}
\end{equation}
Namely, $\sum_{ i = 1 }^N \Delta_{\rho}^2 (A_i) \geq \rm{LB1}$. $\Box$

Theorem \ref{Thm1} provides a series of uncertainty relations depending on the values of the parameter $\alpha$. We can obtain more stringent bound on the uncertainty relations by selecting the optimal parameters $\alpha$. The uncertainty relations (\ref{Zhang_Eq}) is a special case of Theorem \ref{Thm1} corresponding to $\alpha=1$. Note that the lower bound of Theorem \ref{Thm1} is non-zero unless the measured state $|\psi\rangle$ is a common eigenvector of all $A_i$. That is to say, no matter whether the observables are commutable or not, the lower bound of Theorem \ref{Thm1} does not vanish if $|\psi\rangle$ is not a common eigenvector of all observables.

In fact, the lower bound in Theorem 1 should be understood under the permutation of the observables. Let $\pi\in S(N)$ be an arbitrary $N$-element permutation. Define
\begin{equation}
\begin{aligned}
{\rm LB1}_\pi=\mathop{ \max_{ { x \in \{ 0, 1\} } \atop { y \in \{ 0, 1\} } } }
&\frac{1}{ (1 + \alpha^2) (N - 1) }  \bigg\{ \frac{2} { N (N -1) } \Big [ \sum_{ 1 \leq i < j \leq N } \Delta_{\rho} ( \alpha^{1 - x} A_{\pi(i)}  + (- 1)^y \alpha^x A_{\pi(j)} ) \Big ]^2   \\
&  + \sum_{ 1 \leq i < j \leq N } \Delta_{\rho}^2 ( \alpha^ x A_{\pi(i)}  + (- 1)^{1 - y} \alpha^{1 - x} A_{\pi(j)} ) \bigg\}.
\end{aligned}
\end{equation}
The following variance-based uncertainty relation under the element permutation of all observables holds,
\begin{equation}
\sum_{i=1}^N \Delta_{\rho}^2 (A_i) \geq \max_{\pi\in S(N)}{\rm LB1}_\pi.
\end{equation}

According to the following equalities,
\begin{equation}
\sum_{ 1 \leq i < j \leq N } \| a_i + a_j \|^2 =   \| \sum_{i = 1}^N a_i  \|^2 + (N - 2) \sum_{i = 1}^N \| a_i \|^2
\end{equation}
and
\begin{equation}
\sum_{ 1 \leq i < j \leq N } \| a_i - a_j \|^2 =  N \sum_{i = 1}^N \| a_i \|^2 -  \| \sum_{i = 1}^N a_i  \|^2,
\end{equation}
one has \cite{li2022metric},
\begin{equation}\label{li2022}
\begin{aligned}
 \left[\alpha N+(N-2)\beta\right] \sum_{i = 1}^N \|a_i\|^2=
&\beta \sum_{ 1 \leq i < j \leq N } \| a_i + a_j \|^2+ \alpha \sum_{ 1 \leq i < j \leq N } \| a_i - a_j \|^2   \\
&+ ( \alpha - \beta )  \| \sum_{i=1}^N a_i  \|^2 ,
\end{aligned}
\end{equation}
where both $\alpha,\beta$ are arbitrary real numbers.

\begin{theorem}\label{Thm2}
Let $A_1, A_2, \dots, A_N$ be $N$ arbitrary observables. For any quantum state $\rho$, we have the following uncertainty relation satisfied by quantum variances,
\begin{equation}
\sum_{i=1}^N \Delta_{\rho}^2(A_i) \geq \rm{LB2}=\max \{ \rm{X}, \rm{Y}, \rm{Z} \},
\end{equation}
where
\begin{equation}\label{Thm2_Eq1}
\begin{aligned}
\rm{X} =
&\frac{1}{ \alpha N + ( N - 2 ) \beta }  \bigg\{ \frac{ 2\beta }{ N ( N - 1 ) }  \Big [ \sum_{ 1 \leq i < j \leq N } \Delta_{\rho} ( A_i + A_j )  \Big ]^2  \\
&  + \alpha \sum_{ 1 \leq i < j \leq N } \Delta_{\rho}^2 ( A_i - A_j ) + ( \alpha - \beta ) \Delta_{\rho}^2 ( \sum_{ i =1}^{ N } A_i )  \bigg\}
\end{aligned}
\end{equation}
and
\begin{equation}\label{Thm2_Eq2}
\begin{aligned}
\rm{Y} =
&\frac{1}{ \alpha N + ( N - 2 ) \beta }  \bigg\{ \frac{ 2\alpha }{ N ( N - 1 ) } \Big [ \sum_{ 1 \leq i < j \leq N } \Delta_{\rho} ( A_i - A_j ) \Big ]^2  \\
&  + \beta \sum_{ 1 \leq i < j \leq N } \Delta_{\rho}^2 ( A_i + A_j ) + ( \alpha - \beta ) \Delta_{\rho}^2 ( \sum_{ i =1}^{ N } A_i )  \bigg\}
\end{aligned}
\end{equation}
for $ \alpha,\beta  > 0$,
\begin{equation}\label{Thm2_Eq3}
\begin{aligned}
\rm{Z} =
&\frac{1}{ \alpha N + ( N - 2 ) \beta }  \bigg\{ \beta \sum_{ 1 \leq i < j \leq N } \Delta_{\rho}^2 ( A_i + A_j )   \\
&  + \alpha \sum_{ 1 \leq i < j \leq N } \Delta_{\rho}^2 ( A_i - A_j )
 + \frac{ \alpha - \beta }{ ( N - 1 )^2 } \Big [\sum_{ 1 \leq i < j \leq N } \Delta_{\rho} ( A_i + A_j ) \Big ]^2  \bigg\}
\end{aligned}
\end{equation}
for $\beta > \alpha > 0$.
\end{theorem}

{\sf [Proof]} For all $ \alpha,\beta  >0$, by using (\ref{li2022}) and the Cauchy-Schwarz inequality (\ref{cauchy}), we get
\begin{equation}\label{Thm2_Pf_Eq1}
\begin{aligned}
\sum_{i=1}^N \|a_i\|^2 \geq
&\frac{1}{ \alpha N + (N - 2)\beta}  \Big [ \frac{2\beta}{N(N - 1)} ( \sum_{1 \leq i < j \leq N} \|a_i + a_j \| )^2  \\
& + \alpha\sum_{1 \leq i < j \leq N} \|a_i - a_j\|^2 + (\alpha - \beta)  \| \sum_{i = 1}^N a_i \|^2  \Big ]
\end{aligned}
\end{equation}
and
\begin{equation}\label{Thm2_Pf_Eq2}
\begin{aligned}
\sum_{i=1}^N \|a_i\|^2 \geq
&\frac{1}{ \alpha N + (N - 2)\beta}  \Big [ \beta \sum_{1 \leq i < j \leq N} \|a_i + a_j \|^2  \\
& + \frac{2\alpha}{N(N - 1)} ( \sum_{1 \leq i < j \leq N} \|a_i - a_j\| )^2 + (\alpha - \beta)  \| \sum_{i = 1}^N a_i \|^2  \Big ].
\end{aligned}
\end{equation}
When $\beta > \alpha > 0$, due to $ \|\sum\limits_{i = 1}^N a_i \|^2 \leq \frac{1}{(N - 1)^2} ( \sum\limits_{1 \leq i < j \leq N} \|a_i + a_j \| )^2$, we obtain
\begin{equation}\label{Thm2_Pf_Eq3}
\begin{aligned}
\sum_{i=1}^N \|a_i\|^2 \geq
&\frac{1}{\alpha N + (N - 2) \beta}  \Big [ \beta \sum_{1 \leq i < j \leq N} \|a_i + a_j\|^2 + \alpha \sum_{1 \leq i < j \leq N} \|a_i - a_j\|^2  \\
& + \frac{ (\alpha - \beta)}{(N - 1)^2} ( \sum_{1 \leq i < j \leq N} \|a_i + a_j\| )^2  \Big ].
\end{aligned}
\end{equation}
Substituting $\|a_i\| = \Delta_{\rho}(A_i)$ and $\|a_i \pm a_j\| = \Delta_{\rho}(A_i \pm A_j)$ into the inequalities (\ref{Thm2_Pf_Eq1})-(\ref{Thm2_Pf_Eq3}), we complete the proof. $\Box$

In Theorem 2 we note that, for a given amount of observables $N$, the larger $\alpha$ and the smaller $\beta$ mean larger $X$ and $Z$ given by (\ref{Thm2_Eq1}) and (\ref{Thm2_Eq3}), respectively. Nevertheless, the larger $\beta$ and the smaller $\alpha$ correspond to larger $Y$ given by (\ref{Thm2_Eq2}).
If one takes $\alpha=\beta$ in Theorem \ref{Thm2}, the lower bound of Theorem \ref{Thm2} is coincident to that of (\ref{Zhang_Eq}). If one respectively takes $\beta<\alpha$ for {\rm X} and $\beta>\alpha$ for {\rm Y}, the lower bound of Theorem \ref{Thm2} is tighter than that of (\ref{Zhang_Eq}). In \cite{li2022metric}, Li $et\ al.$ proved that (\ref{Thm2_Pf_Eq1}) and (\ref{Thm2_Pf_Eq2}) are strictly tighter than those of norm inequalities related to (\ref{Song_Eq}) and (\ref{Zhang_Eq}) for appropriate $\alpha$ and $\beta$.

In particular, when one takes $\alpha = 2$ and $\beta = 1$ for (\ref{Thm2_Eq1}), $\alpha = 1$ and $\beta = 2$ for (\ref{Thm2_Eq2}) and (\ref{Thm2_Eq3}), then $\rm{X}$, $\rm{Y}$ and $\rm{Z}$ respectively reduce to
\begin{equation}
\begin{aligned}
\rm{X} = &\frac{1}{ 3 N - 2 }  \bigg\{ \frac{ 2 }{ N ( N - 1 ) } \Big [ \sum_{ 1 \leq i < j \leq N } \Delta_{\rho} ( A_i + A_j ) \Big ]^2   \\
&   + 2 \sum_{ 1 \leq i < j \leq N } \Delta_{\rho}^2 ( A_i - A_j ) + \Delta_{\rho}^2 ( \sum_{i = 1}^N A_i )  \bigg\},
\end{aligned}
\end{equation}
\begin{equation}
\begin{aligned}
\rm{Y} =& \frac{1}{ 3 N - 4 }  \bigg\{ \frac{ 2 }{ N ( N - 1 ) } \Big [ \sum_{ 1 \leq i < j \leq N } \Delta_{\rho} ( A_i - A_j ) \Big ]^2    \\
&  + 2 \sum_{ 1 \leq i < j \leq N } \Delta_{\rho}^2 ( A_i + A_j ) - \Delta_{\rho}^2 ( \sum_{i = 1}^N A_i )  \bigg\},
\end{aligned}
\end{equation}
\begin{equation}
\begin{aligned}
\rm{Z} = &\frac{1}{ 3 N - 4 }  \bigg\{ 2 \sum_{ 1 \leq i < j \leq N } \Delta_{\rho}^2 ( A_i + A_j ) + \sum_{ 1 \leq i < j \leq N } \Delta_{\rho}^2 ( A_i - A_j )    \\
&  - \frac{ 1 }{ ( N - 1 )^2 } \Big [\sum_{ 1 \leq i < j \leq N } \Delta_{\rho} ( A_i + A_j ) \Big ]^2  \bigg\}.
\end{aligned}
\end{equation}
For convenience, we consider the above special scenario of Theorem \ref{Thm2} in following concrete examples. We compare the lower bounds $\rm{LB1}$ and $\rm{LB2}$ respectively given in Theorem \ref{Thm1} and \ref{Thm2} with the ones given in (\ref{Song_Eq}) and (\ref{Zhang_Eq}).

{\emph{Example 1}} Consider the qubit mixed state given by Bloch vector $\vec{r} = (\frac{ \sqrt{3} }{2} \cos\theta, \frac{ \sqrt{3} }{2} \sin\theta, 0)$,
\begin{equation}
\rho = \frac{1}{2}(I_2 + \vec{r} \cdot \vec{\sigma}),
\end{equation}
where the components of the vector $\vec{\sigma}=(\sigma_x,\sigma_y,\sigma_z)$ are the standard Pauli matrices, $I_2$ is $2 \times 2$ identity matrix. We choose Pauli matrices $\sigma_x-\sigma_z$, $\sigma_y+\sigma_z$ and $\sigma_z$ as the observables $A_1$, $A_2$ and $A_3$, respectively.

\begin{figure}[tbp]
 \centering
 \subfigure[]
 {
 \label{wzfEx1_3D_Fig.pdf} 
 \includegraphics[width=7.8cm]{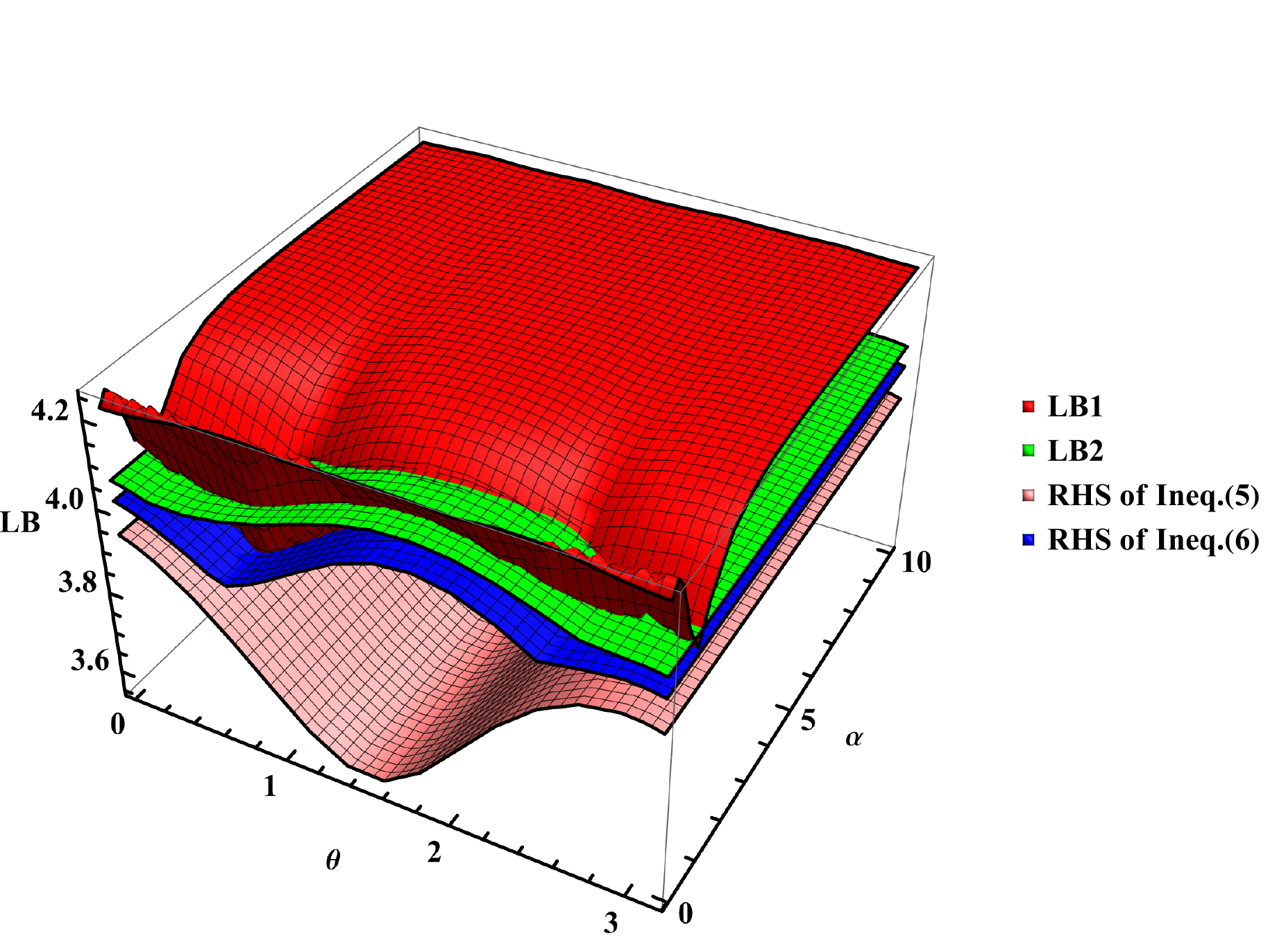}
 }
 \subfigure[]
 {
 \label{wzfEx1_Fig.pdf} 
 \includegraphics[width=7.8cm]{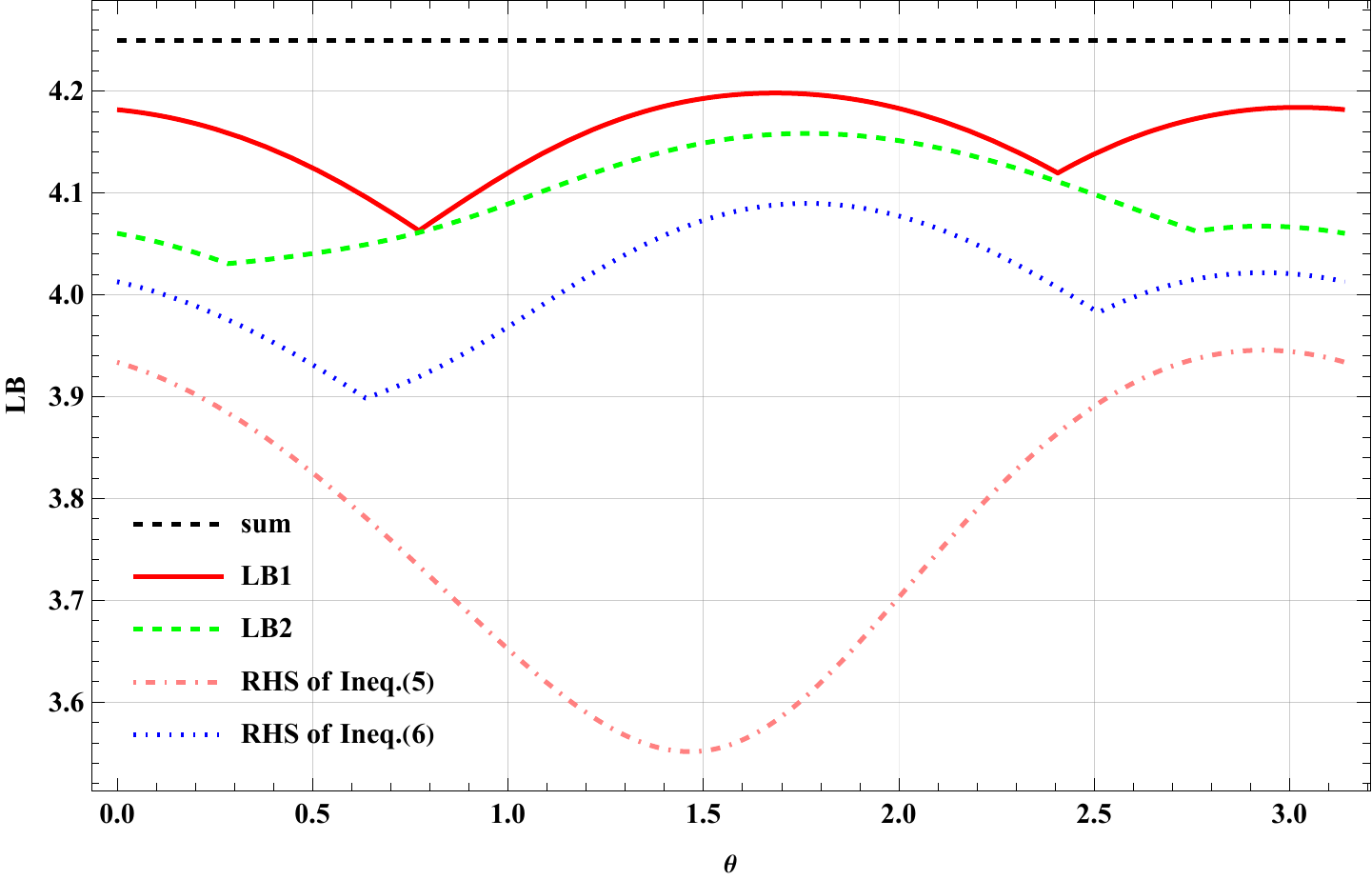}}
  \subfigure[]
 {
 \label{wzfEx1_Fig0.pdf} 
 \includegraphics[width=7.8cm]{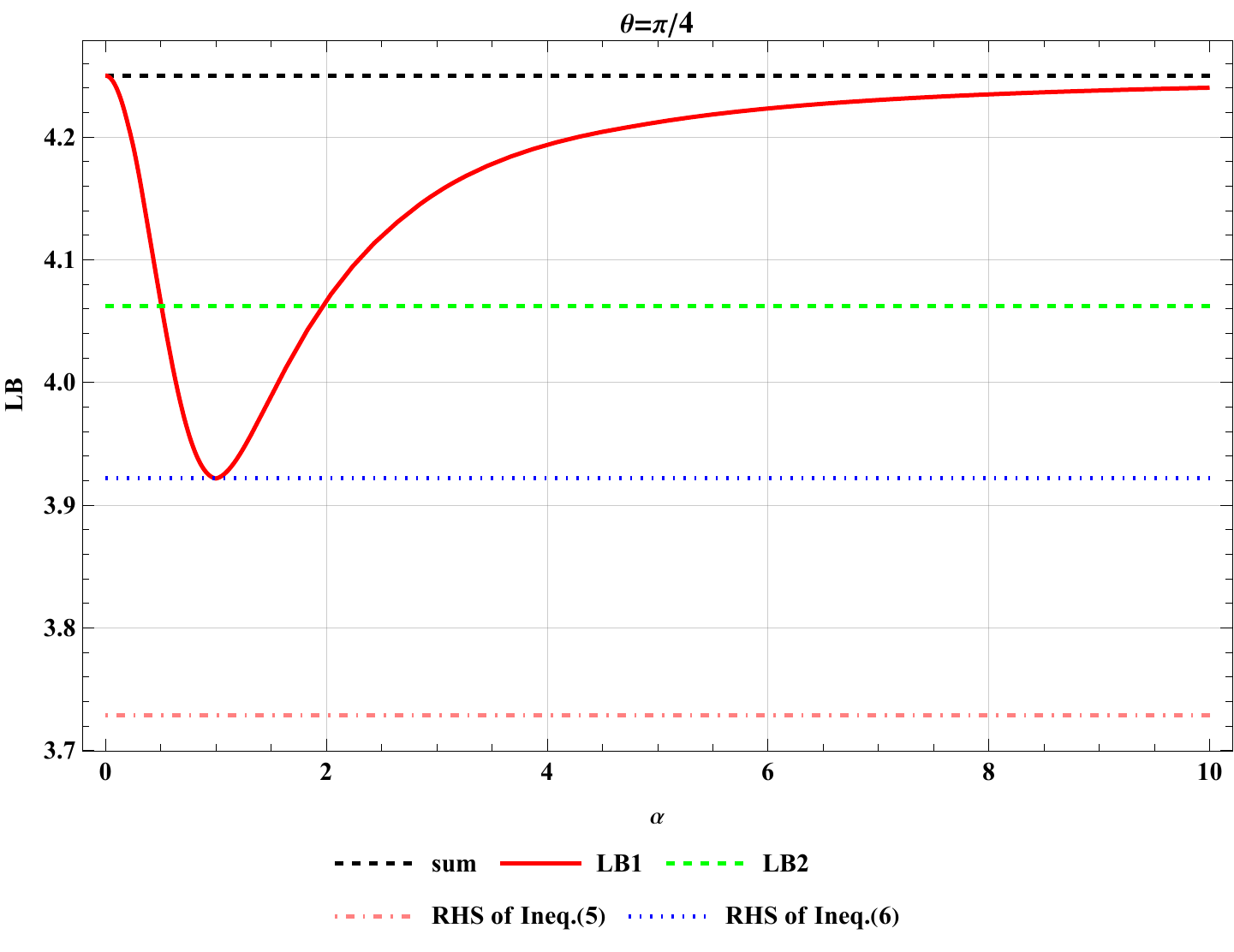}
 }
 \subfigure[]
 {
 \label{wzfEx1_Fig1.pdf} 
 \includegraphics[width=7.8cm]{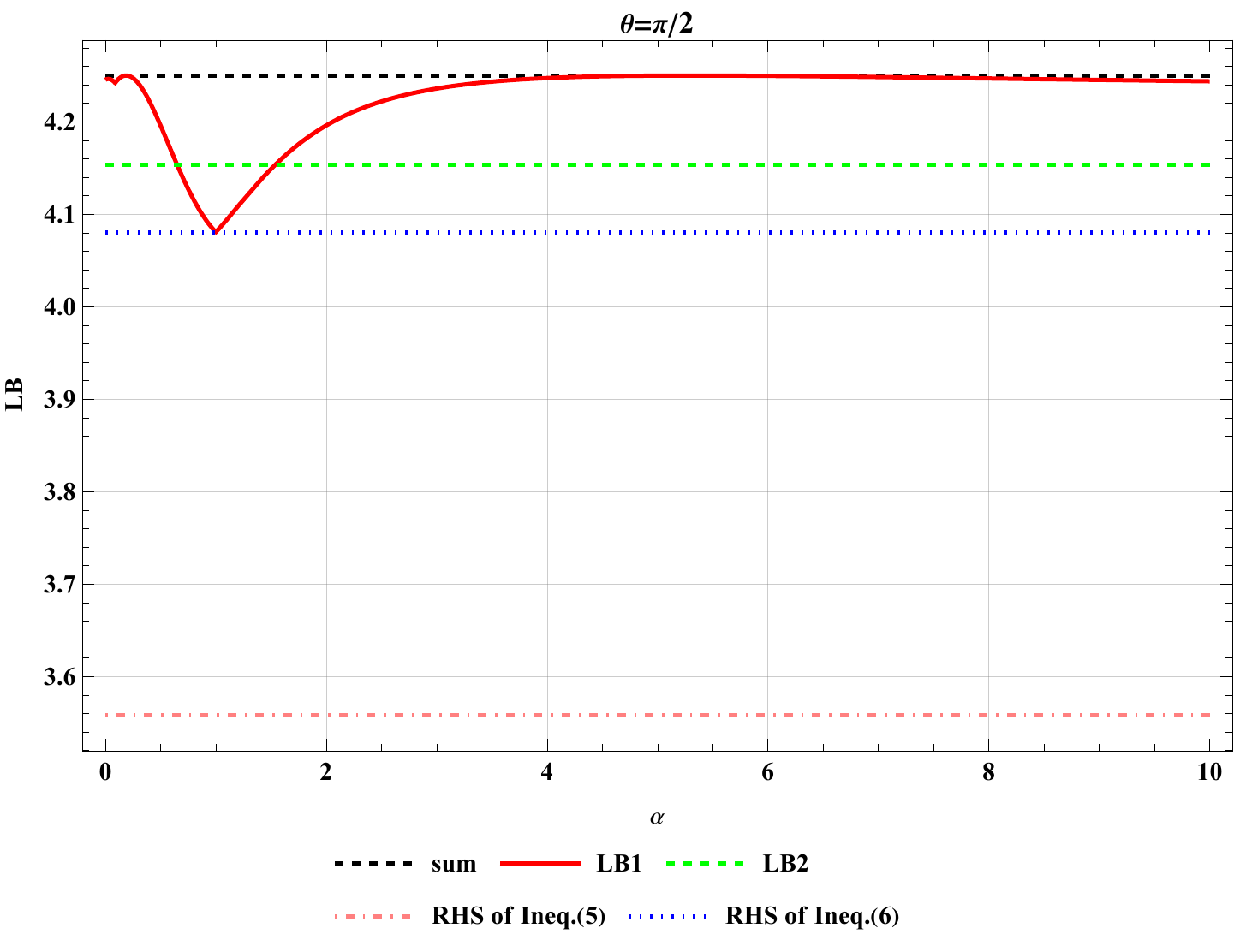}
 }
 \caption{
Example of comparison of \rm{LB1} , \rm{LB2}, the right-hand sides (RHS) of (\ref{Song_Eq}) and (\ref{Zhang_Eq}). We choose three observables $A_1 = \sigma_x-\sigma_z$, $A_2 = \sigma_y+\sigma_z$ and $A_3 = \sigma_z$, and a family of states parameterized by $\theta$. In (a), the red surface denotes the lower bound(\rm{LB1}) of our new uncertainty relation; green surface represent $\rm{LB2}$; pink surface and blue surface represent the right-hand sides (RHS) of (\ref{Song_Eq}) and (\ref{Zhang_Eq}), respectively. In (b), (c) and (d), the upper black (dashed) denotes the sum of variances $\rm{sum} = \Delta_{\rho}^2(A_1) + \Delta_{\rho}^2(A_2) + \Delta_{\rho}^2(A_3)$. The red (solid) curve exhibit $\rm{LB1}$ in (b: $\alpha = 1/2$) and in (c: $\theta = \pi/4$) and (d: $\theta = \pi/2$). The green (dashed) curve represents $\rm{LB2}$. The pink (dot-dashed) and the blue (dotted) curves represent the right-hand sides (RHS) of (\ref{Song_Eq}) and (\ref{Zhang_Eq}), respectively. Note that the new uncertainty relation (\ref{Thm1_Eq1}) is stronger than the others when we select the appropriate parameters.}
 \label{Ex1_Fig}
 \end{figure}

As shown in Fig.~\ref{wzfEx1_3D_Fig.pdf}, we can obtain more stringent bound on the uncertainty relations in most choices of parameter $\alpha$. Especially, set $\alpha = 1/2$ in Theorem \ref{Thm1}, the results are shown in Fig.~\ref{wzfEx1_Fig.pdf}. Obviously, the lower bounds \rm{LB1} and \rm{LB2} are strictly tighter than the bounds of (\ref{Song_Eq}) and (\ref{Zhang_Eq}) in this case. In fact, our lower bonds depend on both the parameter $\alpha$ and the given set of observables. Generally it is difficult to find an analytical relation among the optimal lower bonds, the parameter $\alpha$ and the arbitrary set of observables. To illustrate their relationships, we consider two fixed states with $\theta=\pi/4$ and $\theta=\pi/2$ to show how our lower bounds change with the $\alpha$, see Fig.~\ref{wzfEx1_Fig0.pdf} and Fig.~\ref{wzfEx1_Fig1.pdf}.


{\emph{Example 2} Consider the following class of quantum states given by convex combination of the maximally entangled state and the maximally mixed state,
\begin{equation}\label{isotropic}
\rho_{\theta} = \frac{ 1 - \theta }{ d^2 - 1 } (I_{d^2} - |\Psi^+\ra\la \Psi^+|) + \theta |\Psi^+\ra\la \Psi^+|,
\end{equation}
with $0 \leq \theta \leq 1$ and $|\Psi^+\ra = \frac{1}{ \sqrt{d} } \sum_{i = 1}^d | i i \ra $. Consider two-qubit case ($d = 2$) and take $\sigma_3 \otimes \sigma_1+\sigma_3 \otimes \sigma_2$, $\sigma_3 \otimes \sigma_2$ and $\sigma_3 \otimes \sigma_3-\sigma_3 \otimes \sigma_2$ as the observables $A_1$, $A_2$ and $A_3$, respectively.

As shown in Fig.~\ref{wzfEx2_3D_Fig.pdf}, we can obtain more stringent bound on the uncertainty relations by selecting the appropriate parameter $\alpha$. For comparison, set $\alpha = 5$ in Theorem \ref{Thm1}, the results are shown in Fig.~\ref{wzfEx2_Fig.pdf},  the comparison among our bounds, Song $et\ al.$ and Zhang $et\ al.$'s lower bounds is depicted in Fig.~\ref{Ex2_Fig}.

\begin{figure}[tbp]
 \centering
 \subfigure[]
 {
 \label{wzfEx2_3D_Fig.pdf} 
 \includegraphics[width=7.8cm]{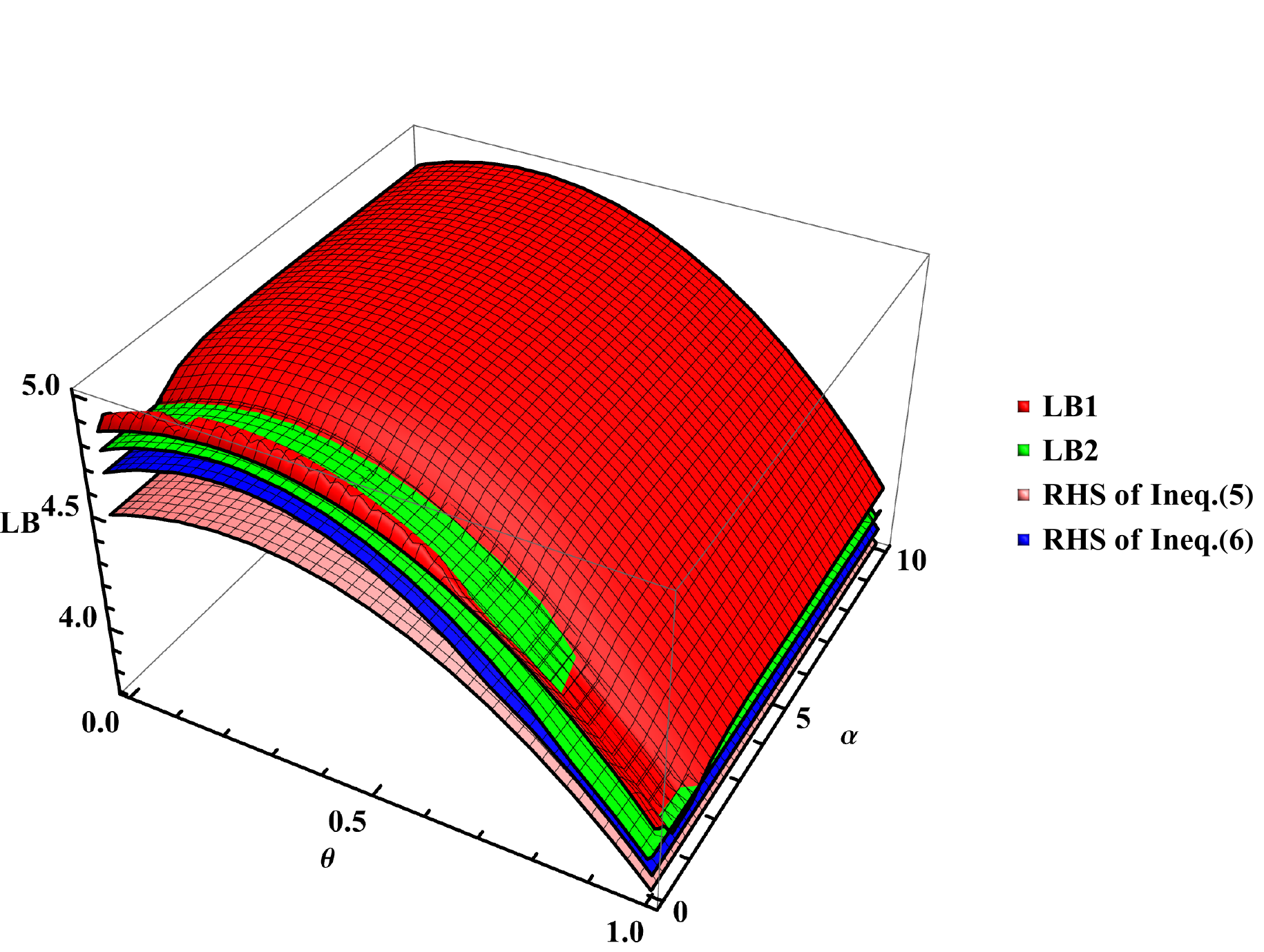}
 }
 \subfigure[]
 {
 \label{wzfEx2_Fig.pdf} 
 \includegraphics[width=7.8cm]{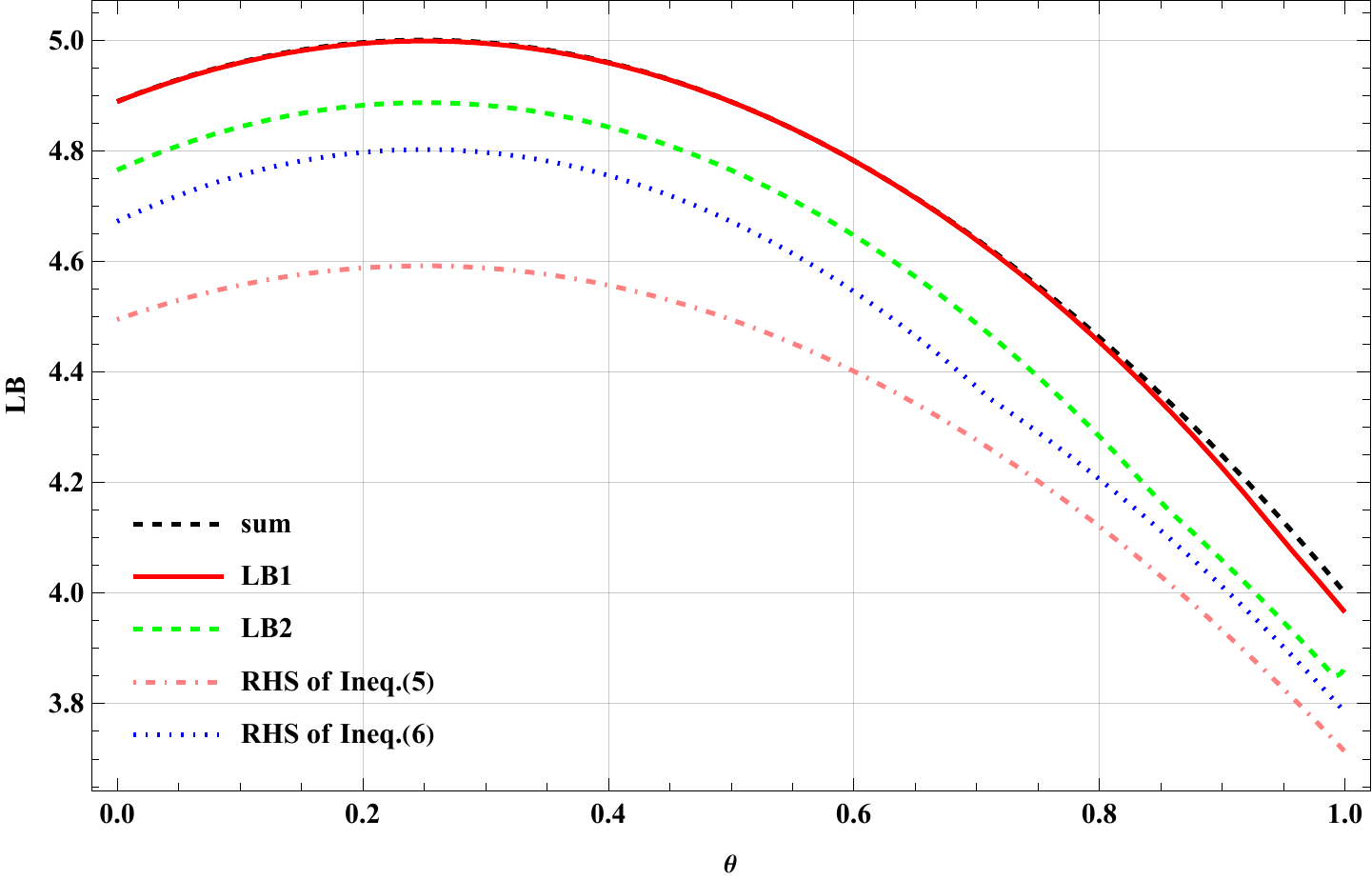}}
 \caption{Comparison among our bounds, Song $et\ al.$ and Zhang $et\ al.$'s bounds for isotropic state (\ref{isotropic}). ({a}) Red and green surfaces respectively represent $\rm{LB1}$ and $\rm{LB2}$. Pink and blue surfaces represent the right-hand sides (RHS) of (\ref{Song_Eq}) and (\ref{Zhang_Eq}), respectively. For most choices of $\alpha$, our lower bounds are able to cover the lower bounds of  (\ref{Song_Eq}) and (\ref{Zhang_Eq}). ({b}) Black (dashed) line is the ${\rm sum} = \Delta_{\rho}^2(A_1) + \Delta_{\rho}^2(A_2) + \Delta_{\rho}^2(A_3)$.  Pink (dot-dashed) and blue (dotted) curves represent the right-hand sides (RHS) of (\ref{Song_Eq}) and (\ref{Zhang_Eq}), respectively. Our bounds \rm{LB1} and \rm{LB2} are shown by the red (solid) and green (dashed) curves, which are larger than ones shown by the blue and pink curves.}
 \label{Ex2_Fig}
 \end{figure}


{\emph{Example 3} Consider the following pure state of spin-1 system,
\begin{equation}
|\psi\ra = \sin\theta \cos\phi |1\ra + \sin\theta \sin\phi |0\ra + \cos\theta|-1\ra,
\end{equation}
where $\theta \in [0, \pi]$ and $\phi \in [0,2\pi]$. We respectively take $L_x-L_y$, $L_y$ and $L_z+L_y$ as the observables $A_1, A_2$ and $A_3$, where $L_x, L_y$ and $L_z$ are the angular momentum operators ($\hbar = 1$):
\begin{equation}
L_x = \frac{1}{ \sqrt{2} }
\begin{pmatrix}
  0 & 1 & 0 \\
  1 & 0 & 1 \\
  0 & 1 & 0
\end{pmatrix},~~~
L_y = \frac{1}{ \sqrt{2} }
\begin{pmatrix}
  0 & -i & 0 \\
  i & 0 & -i \\
  0 & i & 0
\end{pmatrix},~~~
L_z =
\begin{pmatrix}
  1 & 0 & 0 \\
  0 & 0 & 0 \\
  0 & 0 & -1
\end{pmatrix}.
\end{equation}
Set $\phi = \frac{\pi}{2}$. We show in Fig.~\ref{Ex3_Fig} the comparison among our lower bounds of Theorem \ref{Thm1} and Theorem \ref{Thm2}, and those of (\ref{Song_Eq}) and (\ref{Zhang_Eq}). In this scenario, it is easily seen that our bounds are tighter than others  by selecting the appropriate parameter $\alpha$.

\begin{figure}[tbp]
 \centering
 \subfigure[]
 {
 \label{fig:subfig:a} 
 \includegraphics[width=7.8cm]{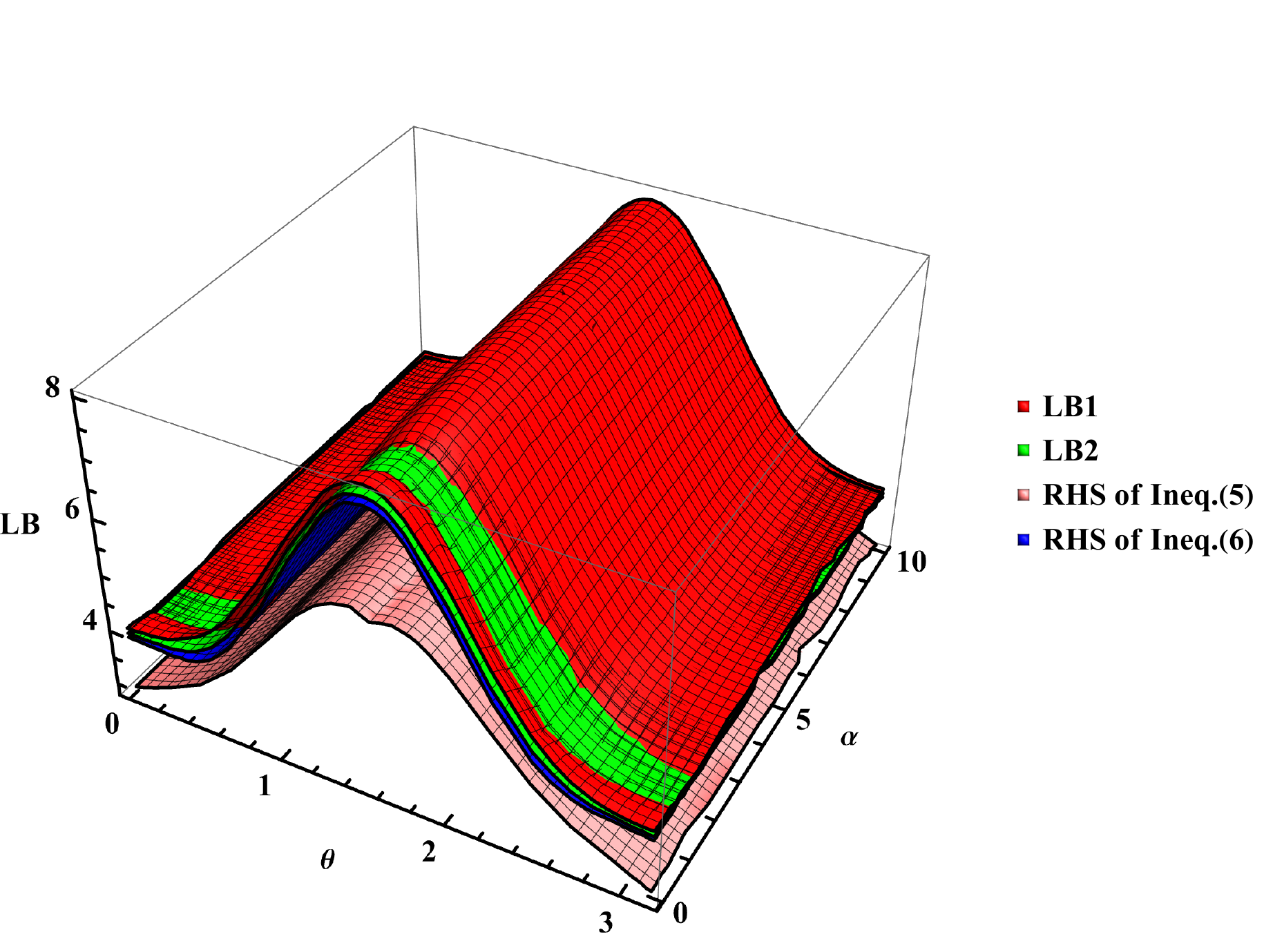}
 }
 \subfigure[]
 {
 \label{fig:subfig:b} 
 \includegraphics[width=7.8cm]{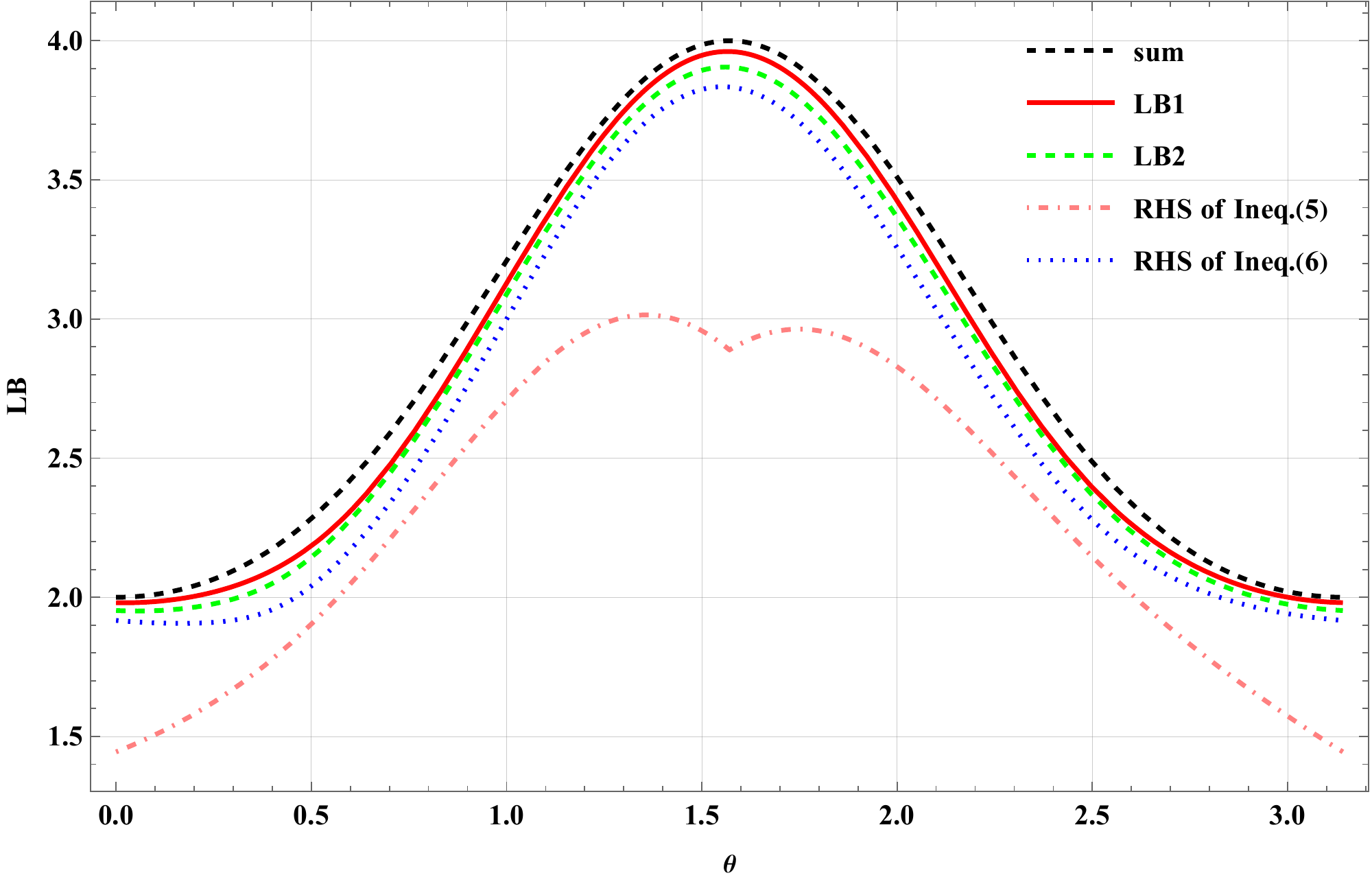}}
 \caption{(\textbf{a}) Red surface represents $\rm{LB1}$, and green surface represents $\rm{LB2}$. Pink and blue surface represent the right-hand sides (RHS) of (\ref{Song_Eq}) and (\ref{Zhang_Eq}), respectively.
 (\textbf{b}) Black (dashed) curve is the ${\rm sum} = \Delta_{\rho}^2(A_1) + \Delta_{\rho}^2(A_2) + \Delta_{\rho}^2(A_3)$. Red (solid) curve represents $\rm{LB1}$ and green (dashed) curve represents $\rm{LB2}$. Pink (dot-dashed) and blue (dotted) curves represent the right-hand sides (RHS) of (\ref{Song_Eq}) and (\ref{Zhang_Eq}), respectively.}
 \label{Ex3_Fig}
 \end{figure}


\section{Conclusion}\label{Sec3}
We have studied tighter variance-based sum uncertainty relations for $N$ arbitrary observables. By employing the parameterized norm identities and Cauchy-Schwarz inequalities we have derived more general and tighter sum uncertainty relations. Furthermore, we have showed that the bounds of our uncertainty relations are tighter than the existing variance-based uncertainty ones. These results and the simple approaches used in this article may highlight further investigations on related uncertainty relations.

\bigskip
\noindent{\bf Acknowledgments}\, This work is supported by NSFC (Grant Nos. 12075159, 12171044), Beijing Natural Science Foundation (Z190005) and the Academician Innovation Platform of Hainan Province.

\noindent{\bf Data availability}\, Data sharing not applicable to this article as no data sets were generated or analyzed during the current study.

\bibliography{Refs}

\begin{thebibliography}{41}
\expandafter\ifx\csname natexlab\endcsname\relax\def\natexlab#1{#1}\fi
\expandafter\ifx\csname bibnamefont\endcsname\relax
  \def\bibnamefont#1{#1}\fi
\expandafter\ifx\csname bibfnamefont\endcsname\relax
  \def\bibfnamefont#1{#1}\fi
\expandafter\ifx\csname citenamefont\endcsname\relax
  \def\citenamefont#1{#1}\fi
\expandafter\ifx\csname url\endcsname\relax
  \def\url#1{\texttt{#1}}\fi
\expandafter\ifx\csname urlprefix\endcsname\relax\def\urlprefix{URL }\fi
\providecommand{\bibinfo}[2]{#2}
\providecommand{\eprint}[2][]{\url{#2}}

\bibitem[{\citenamefont{Heisenberg}(1927)}]{Heisenberg1927}
\bibinfo{author}{\bibfnamefont{W.}~\bibnamefont{Heisenberg}},
  \bibinfo{journal}{Zeitschrift f\"ur Physik} \textbf{\bibinfo{volume}{43}},
  \bibinfo{pages}{198} (\bibinfo{year}{1927}).

\bibitem[{\citenamefont{Nielsen and Chuang}(2002)}]{Nielsen2002Quantum}
\bibinfo{author}{\bibfnamefont{M.~A.} \bibnamefont{Nielsen}} \bibnamefont{and}
  \bibinfo{author}{\bibfnamefont{I.}~\bibnamefont{Chuang}},
  \emph{\bibinfo{title}{Quantum computation and quantum information}}
  (\bibinfo{year}{2002}).

\bibitem[{\citenamefont{Robertson}(1929{\natexlab{a}})}]{PhysRev.34.163}
\bibinfo{author}{\bibfnamefont{H.~P.} \bibnamefont{Robertson}},
  \bibinfo{journal}{Phys. Rev.} \textbf{\bibinfo{volume}{34}},
  \bibinfo{pages}{163} (\bibinfo{year}{1929}{\natexlab{a}}).

\bibitem[{\citenamefont{Schr{\"o}dinger}(1930)}]{schrodinger1930sitzungsberichte}
\bibinfo{author}{\bibfnamefont{E.}~\bibnamefont{Schr{\"o}dinger}},
  \bibinfo{journal}{Acad. Wiss} p. \bibinfo{pages}{296} (\bibinfo{year}{1930}).

\bibitem[{\citenamefont{Maccone and Pati}(2014)}]{PhysRevLett.113.260401}
\bibinfo{author}{\bibfnamefont{L.}~\bibnamefont{Maccone}} \bibnamefont{and}
  \bibinfo{author}{\bibfnamefont{A.~K.} \bibnamefont{Pati}},
  \bibinfo{journal}{Phys. Rev. Lett.} \textbf{\bibinfo{volume}{113}},
  \bibinfo{pages}{260401} (\bibinfo{year}{2014}).

\bibitem[{\citenamefont{Kennard}(1927)}]{Kennard1927Zur}
\bibinfo{author}{\bibfnamefont{E.~H.} \bibnamefont{Kennard}},
  \bibinfo{journal}{Zeitschrift für Physik} \textbf{\bibinfo{volume}{44}},
  \bibinfo{pages}{326} (\bibinfo{year}{1927}).

\bibitem[{\citenamefont{Schr\"odinger}(1930)}]{Schrodinger1930Zum}
\bibinfo{author}{\bibfnamefont{E.}~\bibnamefont{Schr\"odinger}},
  \bibinfo{journal}{Sitzungsberichte der Preussischen Akademie der
  Wissenschaften, Physikalisch-mathematische Klasse}
  \textbf{\bibinfo{volume}{14}}, \bibinfo{pages}{296} (\bibinfo{year}{1930}).

\bibitem[{\citenamefont{Robertson}(1929{\natexlab{b}})}]{Robertson1929The}
\bibinfo{author}{\bibfnamefont{H.~P.} \bibnamefont{Robertson}},
  \bibinfo{journal}{Phys. Rev.} \textbf{\bibinfo{volume}{34}},
  \bibinfo{pages}{163} (\bibinfo{year}{1929}{\natexlab{b}}).

\bibitem[{\citenamefont{Mondal et~al.}(2017)\citenamefont{Mondal, Bagchi, and
  Pati}}]{mondal2017tighter}
\bibinfo{author}{\bibfnamefont{D.}~\bibnamefont{Mondal}},
  \bibinfo{author}{\bibfnamefont{S.}~\bibnamefont{Bagchi}}, \bibnamefont{and}
  \bibinfo{author}{\bibfnamefont{A.~K.} \bibnamefont{Pati}},
  \bibinfo{journal}{Physical Review A} \textbf{\bibinfo{volume}{95}},
  \bibinfo{pages}{052117} (\bibinfo{year}{2017}).

\bibitem[{\citenamefont{Chiew and Gessner}(2022)}]{PhysRevResearch.4.013076}
\bibinfo{author}{\bibfnamefont{S.-H.} \bibnamefont{Chiew}} \bibnamefont{and}
  \bibinfo{author}{\bibfnamefont{M.}~\bibnamefont{Gessner}},
  \bibinfo{journal}{Phys. Rev. Research} \textbf{\bibinfo{volume}{4}},
  \bibinfo{pages}{013076} (\bibinfo{year}{2022}).

\bibitem[{\citenamefont{T\'oth and Fr\"owis}(2022)}]{PhysRevResearch.4.013075}
\bibinfo{author}{\bibfnamefont{G.}~\bibnamefont{T\'oth}} \bibnamefont{and}
  \bibinfo{author}{\bibfnamefont{F.}~\bibnamefont{Fr\"owis}},
  \bibinfo{journal}{Phys. Rev. Research} \textbf{\bibinfo{volume}{4}},
  \bibinfo{pages}{013075} (\bibinfo{year}{2022}).

\bibitem[{\citenamefont{Maassen and
  Uffink}(1988)}]{Maassen1988PhysRevLett.60.1103}
\bibinfo{author}{\bibfnamefont{H.}~\bibnamefont{Maassen}} \bibnamefont{and}
  \bibinfo{author}{\bibfnamefont{J.~B.~M.} \bibnamefont{Uffink}},
  \bibinfo{journal}{Phys. Rev. Lett.} \textbf{\bibinfo{volume}{60}},
  \bibinfo{pages}{1103} (\bibinfo{year}{1988}).

\bibitem[{\citenamefont{Wu et~al.}(2009)\citenamefont{Wu, Yu, and
  M\o{}lmer}}]{Wu2009PhysRevA.79.022104}
\bibinfo{author}{\bibfnamefont{S.}~\bibnamefont{Wu}},
  \bibinfo{author}{\bibfnamefont{S.}~\bibnamefont{Yu}}, \bibnamefont{and}
  \bibinfo{author}{\bibfnamefont{K.}~\bibnamefont{M\o{}lmer}},
  \bibinfo{journal}{Phys. Rev. A} \textbf{\bibinfo{volume}{79}},
  \bibinfo{pages}{022104} (\bibinfo{year}{2009}).

\bibitem[{\citenamefont{Coles et~al.}(2017)\citenamefont{Coles, Berta,
  Tomamichel, and Wehner}}]{Coles2017RevModPhys.89.015002}
\bibinfo{author}{\bibfnamefont{P.~J.} \bibnamefont{Coles}},
  \bibinfo{author}{\bibfnamefont{M.}~\bibnamefont{Berta}},
  \bibinfo{author}{\bibfnamefont{M.}~\bibnamefont{Tomamichel}},
  \bibnamefont{and} \bibinfo{author}{\bibfnamefont{S.}~\bibnamefont{Wehner}},
  \bibinfo{journal}{Rev. Mod. Phys.} \textbf{\bibinfo{volume}{89}},
  \bibinfo{pages}{015002} (\bibinfo{year}{2017}).

\bibitem[{\citenamefont{Busch et~al.}(2013)\citenamefont{Busch, Lahti, and
  Werner}}]{buschPhysRevLett.111.160405}
\bibinfo{author}{\bibfnamefont{P.}~\bibnamefont{Busch}},
  \bibinfo{author}{\bibfnamefont{P.}~\bibnamefont{Lahti}}, \bibnamefont{and}
  \bibinfo{author}{\bibfnamefont{R.~F.} \bibnamefont{Werner}},
  \bibinfo{journal}{Phys. Rev. Lett.} \textbf{\bibinfo{volume}{111}},
  \bibinfo{pages}{160405} (\bibinfo{year}{2013}).

\bibitem[{\citenamefont{Deutsch}(1983)}]{DeutschPhysRevLett.50.631}
\bibinfo{author}{\bibfnamefont{D.}~\bibnamefont{Deutsch}},
  \bibinfo{journal}{Phys. Rev. Lett.} \textbf{\bibinfo{volume}{50}},
  \bibinfo{pages}{631} (\bibinfo{year}{1983}).

\bibitem[{\citenamefont{Distler and Paban}(2013)}]{DistlerPhysRevA.87.062112}
\bibinfo{author}{\bibfnamefont{J.}~\bibnamefont{Distler}} \bibnamefont{and}
  \bibinfo{author}{\bibfnamefont{S.}~\bibnamefont{Paban}},
  \bibinfo{journal}{Phys. Rev. A} \textbf{\bibinfo{volume}{87}},
  \bibinfo{pages}{062112} (\bibinfo{year}{2013}).

\bibitem[{\citenamefont{Pucha{\l}a et~al.}(2013)\citenamefont{Pucha{\l}a,
  Rudnicki, and {\.{Z}}yczkowski}}]{Pucha_a_2013}
\bibinfo{author}{\bibfnamefont{Z.}~\bibnamefont{Pucha{\l}a}},
  \bibinfo{author}{\bibfnamefont{{\L}.}~\bibnamefont{Rudnicki}},
  \bibnamefont{and}
  \bibinfo{author}{\bibfnamefont{K.}~\bibnamefont{{\.{Z}}yczkowski}},
  \bibinfo{journal}{Journal of Physics A: Mathematical and Theoretical}
  \textbf{\bibinfo{volume}{46}}, \bibinfo{pages}{272002}
  (\bibinfo{year}{2013}).

\bibitem[{\citenamefont{Friedland et~al.}(2013)\citenamefont{Friedland,
  Gheorghiu, and Gour}}]{friedland2013universal}
\bibinfo{author}{\bibfnamefont{S.}~\bibnamefont{Friedland}},
  \bibinfo{author}{\bibfnamefont{V.}~\bibnamefont{Gheorghiu}},
  \bibnamefont{and} \bibinfo{author}{\bibfnamefont{G.}~\bibnamefont{Gour}},
  \bibinfo{journal}{Physical review letters} \textbf{\bibinfo{volume}{111}},
  \bibinfo{pages}{230401} (\bibinfo{year}{2013}).

\bibitem[{\citenamefont{Luo}(2003)}]{luoPhysRevLett.91.180403}
\bibinfo{author}{\bibfnamefont{S.}~\bibnamefont{Luo}}, \bibinfo{journal}{Phys.
  Rev. Lett.} \textbf{\bibinfo{volume}{91}}, \bibinfo{pages}{180403}
  (\bibinfo{year}{2003}).

\bibitem[{\citenamefont{Zhang et~al.}(2021{\natexlab{a}})\citenamefont{Zhang,
  Wu, and Fei}}]{Zhang_2021note}
\bibinfo{author}{\bibfnamefont{Q.-H.} \bibnamefont{Zhang}},
  \bibinfo{author}{\bibfnamefont{J.-F.} \bibnamefont{Wu}}, \bibnamefont{and}
  \bibinfo{author}{\bibfnamefont{S.-M.} \bibnamefont{Fei}},
  \bibinfo{journal}{Laser Physics Letters} \textbf{\bibinfo{volume}{18}},
  \bibinfo{pages}{095204} (\bibinfo{year}{2021}{\natexlab{a}}).

\bibitem[{\citenamefont{Zhang and Fei}(2021)}]{zhang2021tighter}
\bibinfo{author}{\bibfnamefont{Q.-H.} \bibnamefont{Zhang}} \bibnamefont{and}
  \bibinfo{author}{\bibfnamefont{S.-M.} \bibnamefont{Fei}},
  \bibinfo{journal}{Quantum Information Processing}
  \textbf{\bibinfo{volume}{20}}, \bibinfo{pages}{1} (\bibinfo{year}{2021}).

\bibitem[{\citenamefont{Ma et~al.}(2022)\citenamefont{Ma, Zhang, and
  Fei}}]{ma2022product}
\bibinfo{author}{\bibfnamefont{X.}~\bibnamefont{Ma}},
  \bibinfo{author}{\bibfnamefont{Q.-H.} \bibnamefont{Zhang}}, \bibnamefont{and}
  \bibinfo{author}{\bibfnamefont{S.-M.} \bibnamefont{Fei}},
  \bibinfo{journal}{Laser Physics Letters} \textbf{\bibinfo{volume}{19}},
  \bibinfo{pages}{055205} (\bibinfo{year}{2022}).

\bibitem[{\citenamefont{G\"uhne}(2004)}]{PhysRevLett.92.117903}
\bibinfo{author}{\bibfnamefont{O.}~\bibnamefont{G\"uhne}},
  \bibinfo{journal}{Phys. Rev. Lett.} \textbf{\bibinfo{volume}{92}},
  \bibinfo{pages}{117903} (\bibinfo{year}{2004}).

\bibitem[{\citenamefont{Zhang and Fei}(2020)}]{zhang2020sufficient}
\bibinfo{author}{\bibfnamefont{Q.-H.} \bibnamefont{Zhang}} \bibnamefont{and}
  \bibinfo{author}{\bibfnamefont{S.-M.} \bibnamefont{Fei}},
  \bibinfo{journal}{Laser Physics Letters} \textbf{\bibinfo{volume}{17}},
  \bibinfo{pages}{065202} (\bibinfo{year}{2020}).

\bibitem[{\citenamefont{Zhang et~al.}(2021{\natexlab{b}})\citenamefont{Zhang,
  Li, Zhang, Fei, and Wang}}]{zhang2021multipartite}
\bibinfo{author}{\bibfnamefont{J.-B.} \bibnamefont{Zhang}},
  \bibinfo{author}{\bibfnamefont{T.}~\bibnamefont{Li}},
  \bibinfo{author}{\bibfnamefont{Q.-H.} \bibnamefont{Zhang}},
  \bibinfo{author}{\bibfnamefont{S.-M.} \bibnamefont{Fei}}, \bibnamefont{and}
  \bibinfo{author}{\bibfnamefont{Z.-X.} \bibnamefont{Wang}},
  \bibinfo{journal}{Scientific reports} \textbf{\bibinfo{volume}{11}},
  \bibinfo{pages}{1} (\bibinfo{year}{2021}{\natexlab{b}}).

\bibitem[{\citenamefont{Giovannetti et~al.}(2006)\citenamefont{Giovannetti,
  Lloyd, and Maccone}}]{PhysRevLett.96.010401}
\bibinfo{author}{\bibfnamefont{V.}~\bibnamefont{Giovannetti}},
  \bibinfo{author}{\bibfnamefont{S.}~\bibnamefont{Lloyd}}, \bibnamefont{and}
  \bibinfo{author}{\bibfnamefont{L.}~\bibnamefont{Maccone}},
  \bibinfo{journal}{Phys. Rev. Lett.} \textbf{\bibinfo{volume}{96}},
  \bibinfo{pages}{010401} (\bibinfo{year}{2006}).

\bibitem[{\citenamefont{Schneeloch et~al.}(2013)\citenamefont{Schneeloch,
  Broadbent, Walborn, Cavalcanti, and Howell}}]{SchneelochPhysRevA.87.062103}
\bibinfo{author}{\bibfnamefont{J.}~\bibnamefont{Schneeloch}},
  \bibinfo{author}{\bibfnamefont{C.~J.} \bibnamefont{Broadbent}},
  \bibinfo{author}{\bibfnamefont{S.~P.} \bibnamefont{Walborn}},
  \bibinfo{author}{\bibfnamefont{E.~G.} \bibnamefont{Cavalcanti}},
  \bibnamefont{and} \bibinfo{author}{\bibfnamefont{J.~C.}
  \bibnamefont{Howell}}, \bibinfo{journal}{Phys. Rev. A}
  \textbf{\bibinfo{volume}{87}}, \bibinfo{pages}{062103}
  (\bibinfo{year}{2013}).

\bibitem[{\citenamefont{Hall}(2005)}]{hall2005exact}
\bibinfo{author}{\bibfnamefont{M.~J.} \bibnamefont{Hall}},
  \bibinfo{journal}{General Relativity and Gravitation}
  \textbf{\bibinfo{volume}{37}}, \bibinfo{pages}{1505} (\bibinfo{year}{2005}).

\bibitem[{\citenamefont{Fuchs and Peres}(1996)}]{Fuchs1996Quantum}
\bibinfo{author}{\bibfnamefont{C.~A.} \bibnamefont{Fuchs}} \bibnamefont{and}
  \bibinfo{author}{\bibfnamefont{A.}~\bibnamefont{Peres}},
  \bibinfo{journal}{Phys. Rev. A} \textbf{\bibinfo{volume}{53}},
  \bibinfo{pages}{2038} (\bibinfo{year}{1996}).

\bibitem[{\citenamefont{Kechrimparis and
  Weigert}(2014)}]{kechrimparis2014heisenberg}
\bibinfo{author}{\bibfnamefont{S.}~\bibnamefont{Kechrimparis}}
  \bibnamefont{and} \bibinfo{author}{\bibfnamefont{S.}~\bibnamefont{Weigert}},
  \bibinfo{journal}{Physical Review A} \textbf{\bibinfo{volume}{90}},
  \bibinfo{pages}{062118} (\bibinfo{year}{2014}).

\bibitem[{\citenamefont{Dammeier et~al.}(2015)\citenamefont{Dammeier,
  Schwonnek, and Werner}}]{dammeier2015uncertainty}
\bibinfo{author}{\bibfnamefont{L.}~\bibnamefont{Dammeier}},
  \bibinfo{author}{\bibfnamefont{R.}~\bibnamefont{Schwonnek}},
  \bibnamefont{and} \bibinfo{author}{\bibfnamefont{R.~F.}
  \bibnamefont{Werner}}, \bibinfo{journal}{New Journal of Physics}
  \textbf{\bibinfo{volume}{17}}, \bibinfo{pages}{093046}
  (\bibinfo{year}{2015}).

\bibitem[{\citenamefont{Ma et~al.}(2017)\citenamefont{Ma, Chen, Liu, Wang, Ye,
  Kong, Shi, Fei, and Du}}]{ma2017experimental}
\bibinfo{author}{\bibfnamefont{W.}~\bibnamefont{Ma}},
  \bibinfo{author}{\bibfnamefont{B.}~\bibnamefont{Chen}},
  \bibinfo{author}{\bibfnamefont{Y.}~\bibnamefont{Liu}},
  \bibinfo{author}{\bibfnamefont{M.}~\bibnamefont{Wang}},
  \bibinfo{author}{\bibfnamefont{X.}~\bibnamefont{Ye}},
  \bibinfo{author}{\bibfnamefont{F.}~\bibnamefont{Kong}},
  \bibinfo{author}{\bibfnamefont{F.}~\bibnamefont{Shi}},
  \bibinfo{author}{\bibfnamefont{S.-M.} \bibnamefont{Fei}}, \bibnamefont{and}
  \bibinfo{author}{\bibfnamefont{J.}~\bibnamefont{Du}},
  \bibinfo{journal}{Physical Review Letters} \textbf{\bibinfo{volume}{118}},
  \bibinfo{pages}{180402} (\bibinfo{year}{2017}).

\bibitem[{\citenamefont{Chen et~al.}(2016)\citenamefont{Chen, Cao, Fei, and
  Long}}]{chen2016variance}
\bibinfo{author}{\bibfnamefont{B.}~\bibnamefont{Chen}},
  \bibinfo{author}{\bibfnamefont{N.-P.} \bibnamefont{Cao}},
  \bibinfo{author}{\bibfnamefont{S.-M.} \bibnamefont{Fei}}, \bibnamefont{and}
  \bibinfo{author}{\bibfnamefont{G.-L.} \bibnamefont{Long}},
  \bibinfo{journal}{Quantum Information Processing}
  \textbf{\bibinfo{volume}{15}}, \bibinfo{pages}{3909} (\bibinfo{year}{2016}).

\bibitem[{\citenamefont{Chen and Fei}(2015)}]{chen2015sum}
\bibinfo{author}{\bibfnamefont{B.}~\bibnamefont{Chen}} \bibnamefont{and}
  \bibinfo{author}{\bibfnamefont{S.-M.} \bibnamefont{Fei}},
  \bibinfo{journal}{Scientific reports} \textbf{\bibinfo{volume}{5}},
  \bibinfo{pages}{1} (\bibinfo{year}{2015}).

\bibitem[{\citenamefont{Chen et~al.}(2019)\citenamefont{Chen, Wang, Li, Song,
  and Qiao}}]{chen2019tight}
\bibinfo{author}{\bibfnamefont{Z.-X.} \bibnamefont{Chen}},
  \bibinfo{author}{\bibfnamefont{H.}~\bibnamefont{Wang}},
  \bibinfo{author}{\bibfnamefont{J.-L.} \bibnamefont{Li}},
  \bibinfo{author}{\bibfnamefont{Q.-C.} \bibnamefont{Song}}, \bibnamefont{and}
  \bibinfo{author}{\bibfnamefont{C.-F.} \bibnamefont{Qiao}},
  \bibinfo{journal}{Scientific Reports} \textbf{\bibinfo{volume}{9}},
  \bibinfo{pages}{1} (\bibinfo{year}{2019}).

\bibitem[{\citenamefont{Qin et~al.}(2016)\citenamefont{Qin, Fei, and
  Li-Jost}}]{qin2016multi}
\bibinfo{author}{\bibfnamefont{H.-H.} \bibnamefont{Qin}},
  \bibinfo{author}{\bibfnamefont{S.-M.} \bibnamefont{Fei}}, \bibnamefont{and}
  \bibinfo{author}{\bibfnamefont{X.}~\bibnamefont{Li-Jost}},
  \bibinfo{journal}{Scientific Reports} \textbf{\bibinfo{volume}{6}},
  \bibinfo{pages}{1} (\bibinfo{year}{2016}).

\bibitem[{\citenamefont{Xiao and Jing}(2016)}]{xiao2016mutually}
\bibinfo{author}{\bibfnamefont{Y.}~\bibnamefont{Xiao}} \bibnamefont{and}
  \bibinfo{author}{\bibfnamefont{N.}~\bibnamefont{Jing}},
  \bibinfo{journal}{Scientific Reports} \textbf{\bibinfo{volume}{6}},
  \bibinfo{pages}{1} (\bibinfo{year}{2016}).

\bibitem[{\citenamefont{Song et~al.}(2017)\citenamefont{Song, Li, Peng, and
  Qiao}}]{song2017stronger}
\bibinfo{author}{\bibfnamefont{Q.-C.} \bibnamefont{Song}},
  \bibinfo{author}{\bibfnamefont{J.-L.} \bibnamefont{Li}},
  \bibinfo{author}{\bibfnamefont{G.-X.} \bibnamefont{Peng}}, \bibnamefont{and}
  \bibinfo{author}{\bibfnamefont{C.-F.} \bibnamefont{Qiao}},
  \bibinfo{journal}{Scientific Reports} \textbf{\bibinfo{volume}{7}},
  \bibinfo{pages}{1} (\bibinfo{year}{2017}).

\bibitem[{\citenamefont{Zhang et~al.}(2023)\citenamefont{Zhang, Wu, Ma, and
  Fei}}]{zhang2022note}
\bibinfo{author}{\bibfnamefont{Q.-H.} \bibnamefont{Zhang}},
  \bibinfo{author}{\bibfnamefont{J.-F.} \bibnamefont{Wu}},
  \bibinfo{author}{\bibfnamefont{X.}~\bibnamefont{Ma}}, \bibnamefont{and}
  \bibinfo{author}{\bibfnamefont{S.-M.} \bibnamefont{Fei}},
  \bibinfo{journal}{Quantum Information Processing}
  \textbf{\bibinfo{volume}{22}}, \bibinfo{pages}{1} (\bibinfo{year}{2023}).

\bibitem[{\citenamefont{Li et~al.}(2022)\citenamefont{Li, Gao, and
  Yan}}]{li2022metric}
\bibinfo{author}{\bibfnamefont{H.}~\bibnamefont{Li}},
  \bibinfo{author}{\bibfnamefont{T.}~\bibnamefont{Gao}}, \bibnamefont{and}
  \bibinfo{author}{\bibfnamefont{F.}~\bibnamefont{Yan}},
  \bibinfo{journal}{Physica Scripta} \textbf{\bibinfo{volume}{98}},
  \bibinfo{pages}{015024} (\bibinfo{year}{2022}).

\end{thebibliography}
\end{document}